\documentclass[preprint,showpacs,jcp,amsmath,amssymb]{revtex4-1}
\usepackage{graphicx}
\usepackage{epsfig}
\usepackage{bm}
\begin{document}

\title{Mechanism of Molecular Orientation by Single-cycle Pulses}

\author{Juan Ortigoso}

\affiliation{Instituto de Estructura de la Materia, CSIC, Serrano 121, 28006 
Madrid, Spain}

\date{24 July 2012}

\begin{abstract}
Significant molecular orientation can be achieved by time-symmetric single-cycle pulses of zero area, in the THz region. We show that in spite of the existence of a combined time-space symmetry operation, not only large peak instantaneous orientations but also nonzero time-average orientations over a rotational period can be obtained.  We show that this unexpected phenomenon is due to interferences among eigenstates of the time-evolution operator, as was described previously for transport phenomena in quantum ratchets.  This mechanism also works for sequences of identical pulses, spanning a rotational period. This fact can be used to obtain a net average molecular orientation regardless of the magnitude of the rotational constant. 
\end{abstract}

\maketitle

\section{Introduction}\label{uno}

Control of molecular alignment and orientation is one of the goals of current research in chemical physics. In the last two decades, since the demonstration of the so-called brute-force orientation \cite{loesch, friedrich1}, impressive progress has taken place in both areas. The theoretical basis of these phenomena and the basic experimental techniques have been described in thorough reviews \cite{seide1, seide1b}.  

Alignment refers to placing the molecular axis (for linear molecules) parallel to a space fixed axis \cite{seide1}.  A strong linearly polarized laser field, out of resonance with electronic transitions, is able to excite a coherent superposition of rotational eigenstates, giving rise to localized rotational wave packets \cite{bretis2}. The nonresonant laser interacts with molecules via molecular polarizability. Under appropriate conditions, the intermolecular axis becomes localized parallel to the polarization vector of the laser field. 

Orientation requires that the molecular axis has, in addition, a defined direction with respect to the space axis. 
Thus, achievement of molecular orientation by time-dependent fields has been considered traditionally more difficult than the well understood molecular alignment since it requires breakdown of inversion symmetry.  Three basic schemes have been proposed, (i) combination of a strong nonresonant field and a weak static field \cite{japo, bretis, bretism, bretisb, ortisantos, feres}, (ii)  two-color laser schemes with resonant \cite{ twocolor} or nonresonant \cite{twocolor2, twocolor2b} pulses, and (iii) sub-cycle and few cycle THz pulses \cite{chin, fcp}.  Also, control with tailored microwave fields has been proposed \cite{orti1, salomon}.  Further progress is centered in the development of applications \cite{stapel1} and new all-optical techniques \cite{japo2}. 

It is frequently argued that any orientational effect in molecules requires an external field asymmetric in time. Thus, the highly asymmetric temporal structure of half-cycle pulses was considered especially appropriate to achieve orientation \cite{orien1}. 

Single-cycle pulses of zero area, in the THz region, have not been considered especially useful for achieving molecular orientation due to their lack of time asymmetry. In effect, molecular orientation requires, in principle, simultaneous breakdown of all spatio-temporal symmetries. However, a related phenomenon, the generation of 
transient currents for quantum Hamiltonian ratchets can be achieved in the presence of some symmetry \cite{hanggi, santos, santosb, lehman}.  Recently, it has been shown that directed motion in the absence of an external bias (the ratchet effect), can occur in coherent systems such as a Bose-Einstein condensate in an optical lattice periodically modulated in time \cite{salger}. 

As far as we know, only a paper by Sugny {\it et al.} \cite{zero}, and a recent work by Fleischer {\it et al.} \cite{bob}  have studied molecular orientation with pulses of zero area.  Numerical evidence was presented in \cite{zero} showing that a significant instantaneous orientation can be produced during and after the pulse is over. The main purpose of the work by Sugny {\it et al.} \cite{zero} was testing the efficiency of a time-dependent unitary perturbation theory, and therefore details of the wave packets involved in the orientation phenomenon were not analyzed. On the other hand, a fairly complete numerical analysis of the molecule and field parameters needed to produce maximum orientation was given.

Recently, it has been experimentally demonstrated \cite{bob}, for the first time, that intense single-cycle THz pulses induce field-free orientation that survives thermal averaging. The emphasis of the work by Fleischer {\it et al.} \cite{bob} was the experimental observation of the phenomenon. These authors calculated and measured the degree of orientation and alignment that a THz pulse is able to produce in a sample of OCS molecules. 

The aim of the present work is to study the mechanism for which molecules can get oriented with symmetric pulses in the light of previous studies in quantum ratchets. We have found that a common explanation can be given for both phenomena, based on the existence of interferences between pairs of eigenstates of the unitary time propagator. 

Here, by considering  that the orienting pulse is a member of a periodic sequence of identical pulses, we study in Sec. \ref{dos} the symmetries of the Floquet Hamiltonian instead of the more complicated unitary time propagator, as the eigenvectors of both operators are related \cite{yajima}. From this symmetry analysis we show that a nonzero average orientation during a pulse can be achieved. In Sec. \ref{cuatro} we present numerical evidence of this phenomenon, and study its dependence on the duration and strength of the electric field. We show that not only large peak orientations but also a nonzero time average orientation during a rotational period can be obtained.  In Sec. {\ref{cinco} we present our general conclusions. Specifically, we discuss the advantages of using an unique pulse whose duration is comparable to a rotational period or a pulse sequence that span such a time length. Also, we argue that the suitability of each approach depends on the magnitude of the rotational constant. Finally, a brief summary of the Floquet approach is presented in an Appendix.

\section{Symmetry analysis for polar molecules interacting with single-cycle pulses} \label{dos}

\subsection{Symmetry properties of quasienergy eigenfunctions}\label{dosb}

The electromagnetic field for a periodic field formed by single-cycle pulses is an odd function of time, $E(t)=-E(-t)$. Due to this property, the eigenvalue equation for the Floquet Hamiltonian
(see Appendix), ${\cal F}(t')\widehat{\chi}(t')=\epsilon \widehat{\chi}(t')$, is invariant under the operation \cite{lehman}

\begin{equation}
S_{\rm rTP}: (\widehat{\chi}, \theta, t')\rightarrow (\widehat{\chi}^{*}, \pi-\theta, -t') \;.
\label{simetria1}
\end{equation}

\noindent
This anti-linear transformation is the rotational version of the so-called time-inversion parity $S_{\rm TP}$,  which generalizes the notion of parity to the extended Hilbert space 
\cite{hanggi, lehman}. The transformation $\theta\rightarrow \pi-\theta$,  is a rotational-restricted version of the operation $E^{*}$, that in molecules inverts the spatial coordinates of all the nuclei and electrons through the molecular center of mass \cite{bunker}.

Eigenstates $\widehat{\chi}$ can be symmetric (+) or antisymmetric (-) with respect to time-inversion parity, i.e.,

\begin{equation}
\widehat{\chi}(\theta, t')=\pm \widehat{\chi}^{*}(\pi-\theta,-t')\;.
\label{pcccp4}
\end{equation}

\noindent 
This symmetry along with the property

\begin{equation}
E^{*}|J,M\rangle=(-1)^{J+M}|J,M\rangle \;,
\label{pccp5}
\end{equation}

\noindent
implies that the coefficients $d_{Jn}$ in Eq. (\ref{eq2}) are purely imaginary for $J$ odd and real for $J$ even, or viceversa depending, both on the symmetric or antisymmetric character of the Floquet eigenstate and the $M$ value of the initial state. Thus, for symmetric eigenfunctions, the coefficients $d_{Jn}$ are real if $J+M$ is even, and pure imaginary if $J+M$ is odd. On the other hand, the use of these restrictions in the expression, Eq. (\ref{eq4}), gives 

\begin{equation}
c_{J}(t)=(-1)^{(J+M+l)} c_{J}^{*}(-t),\;\; l=0, 1\;,
\label{pccp6}
\end{equation}

\noindent
where $l=0$ for symmetric functions, and $l=1$ for antisymmetric. This relationship between the coefficients at times $t$ and $-t$,
along with the following selection rules for the matrix elements of $\cos \theta$ between rotational eigenstates,

\begin{equation}
\langle J, M|\cos \theta |J',M\rangle \neq 0, {\rm for} \; \Delta J=\pm 1, \Delta M=0 \;,
\label{selectionrule}
\end{equation}

\noindent
gives for the matrix elements between Floquet eigenfunctions, in the standard Hilbert space, the following identities

\begin{equation}
\langle \chi_{n}(t)|\cos \theta| \chi_{n}(t)\rangle=-\langle \chi_{n}(-t)|\cos \theta| \chi_{n}(-t)\rangle \;,
\label{rel1}
\end{equation}

\begin{equation}
\langle \chi_{n}^{\pm}(t)|\cos \theta| \chi_{m}^{\pm}(t)\rangle=-\langle \chi_{m}^{\pm}(-t)|\cos \theta| \chi_{n}^{\pm}(-t)\rangle \;,
\label{rel2}
\end{equation}

\begin{equation}
\langle \chi_{n}^{\pm}(t)|\cos \theta| \chi_{m}^{\mp}(t)\rangle=\langle \chi_{m}^{\mp}(-t)|\cos \theta| \chi_{n}^{\pm}(-t)\rangle \;.
\label{rel22}
\end{equation}

\noindent
Eq. (\ref{rel1}) inmediately implies

\begin{equation}
\int_{-T/2}^{T/2} \langle \chi_{n}(t)| \cos \theta |\chi_{n}(t)\rangle dt=0, \; \forall n.
\label{resul1}
\end{equation}

Note that matrix elements, Eqs. (\ref{rel2}) and (\ref{rel22}), can be complex. Thus,  the integrals $\int_{-T/2}^{T/2} \langle \chi_{n}^{\pm}(t)|\cos \theta|\chi_{m}^{\mp}(t) \rangle dt$
and  $\int_{-T/2}^{T/2} \langle \chi_{n}^{\pm}(t)|\cos \theta|\chi_{m}^{\pm}(t) \rangle dt$ for $n\ne m$ are, in general,  nonzero.  However, linear combinations of functions with the same time-inversion parity  give

\begin{equation}
\int_{-T/2}^{T/2} \langle (b_{n}\chi_{n}^{\pm} + b_{m} \chi_{m}^{\pm})| \cos\theta|( b_{n}\chi_{n}^{\pm}+b_{m} \chi_{m}^{\pm} )\rangle dt= 0,
\label{resul3}
\end{equation}

\noindent
if $b_{n}$, $b_{m}$ are both real or pure imaginary. In the same way, for linear combination of functions with different symmetry, we have

\begin{equation}
\int_{-T/2}^{T/2} \langle (b_{n}\chi_{n}^{\pm} + b_{m} \chi_{m}^{\mp})| \cos\theta|( b_{n}\chi_{n}^{\pm}+b_{m} \chi_{m}^{\mp} )\rangle dt= 0,
\label{resulb3}
\end{equation}

\noindent
if one of the coefficients $b$ is real and the other imaginary. When $b_{n}$ and $b_{m}$ are both complex, the integrals are nonzero.

\subsection{Symmetry properties of time-dependent wave packets}\label{dosc}

The average orientation over the pulse duration, for an arbitrary initial state, $\Psi(-T/2)$, is

\begin{equation}
\langle\langle \cos\theta\rangle \rangle=\frac{1}{T}\int_{-T/2}^{T/2} \sum_{n\ne m}  \exp\left[i \left(\epsilon_{n}-\epsilon_{m}\right)\left(t+T/2\right) \right]b_{n}^{*}b_{m} \langle \chi_{n}(t)| \cos \theta |\chi_{m} (t)\rangle  dt \;.
\label{nuevodesa}
\end{equation}

\noindent
where 

\begin{equation}
b_{n}=\langle \chi_{n}(t=-T/2)|\Psi \rangle \;.
\label{pccp2}
\end{equation}

\noindent
When the initial state is a Floquet eigenstate, $\chi_{n}(-T/2)$,  only one term arises in the summation, Eq. (\ref{eigenpsi}), and from Eq. (\ref{resul1}) and (\ref{nuevodesa}), we get

\begin{equation}
\langle\langle \cos\theta\rangle \rangle=\int_{-T/2}^{mT/2} \langle \Psi(t)|\cos \theta|\Psi(t)\rangle =0 \;,
\label{reresul1}
\end{equation}

\noindent
for any number of pulses $m$. Thus, if a molecule, initially described by a given Floquet eigenstate, is oriented (well localized at $\theta_{0}$) at time $-t$ it becomes antioriented (well localized at 
$\pi-\theta_{0}$) at time $t$, producing a zero time-average orientation during a pulse or sequence of pulses. 

Using relations Eq. (\ref{rel1})-(\ref{rel22}), we can write Eq. (\ref{nuevodesa}) as

\begin{equation}
\langle\langle \cos\theta\rangle \rangle= \frac{1}{T}\int_{0}^{T/2} dt \sum_{m>n} A_{nm} \Theta [\exp [i (\epsilon_{n}-\epsilon_{m}) t]\langle \chi_{n}(t)| \cos \theta |\chi_{m} (t)\rangle],
\label{eqlong}
\end{equation}

\noindent
where the function $\Theta[.]={\rm Re}[.]$ if $l_{n}\ne l_{m}$, and $\Theta[.]=i \times {\rm Im}[.]$ if $l_{n}=l_{m}$, with the constants $A_{nm}$ defined as

\begin{equation}
A_{nm}=2 \{b_{n}^{*}b_{m} \exp[i(\epsilon_{n}-\epsilon_{m})T/2]-(-1)^{(l_{n}+l_{m})} b_{n} b_{m}^{*} \exp[i(\epsilon_{m}-\epsilon_{n}) T/2]\}.
\end{equation}

\noindent
This expression shows that the time average over the pulse duration of $\langle \cos\theta\rangle (t)$ can be obtained integrating exclusively over the second half of the pulse. Since there  is no symmetry relation between $\langle \cos\theta \rangle(t)$ at any two different times greater than zero, the integral in Eq. (\ref{eqlong}) 
can be nonzero in spite of the existence of time-reversal parity. This mechanism can be loosely called transient breaking of time-reversal parity. Thus, symmetry relations do not forbid a time average orientation, during a single pulse. 

From a classical point of view, for each classical trajectory with instantaneous
orientation $\theta(t)$ there exists another one with orientation $\pi-\theta(t)$. Thus, the average orientation in the classical ergodic limit is zero \cite{hanggi}.
Although time-reversal parity does not forbid a net time average orientation for an arbitrary initial state $\Psi(t)$ during a single pulse, only nondiagonal matrix elements between Floquet eigenstates contribute to the average, and, in the long-time limit, the quantum time average goes to zero too. In effect, for a sequence of $l_{\rm max}$ pulses, with repetition period $T$, the average orientation for an arbitrary initial state $\Psi$, given by Eq. (\ref{eigenpsi}), is

\begin{equation}
\langle\langle \cos \theta \rangle \rangle_{Tl_{\rm max}}=\frac{1}{(l_{\rm max}+1)T} \sum_{l=0}^{l_{\rm max}} \int_{-T/2+lT}^{-T/2+(l+1)T} \langle \Psi(t)|\cos\theta|\Psi(t)\rangle dt \;,
\label{promediopre}
\end{equation}

\noindent
which can be written, thanks to the periodicity of the $\chi$ functions \cite{santosb}, as

\begin{equation}
\langle\langle \cos \theta \rangle \rangle_{T l_{\rm max}}=
\sum_{nm}b_{n}^{*}b_{m}e^{i (\epsilon_{n}-\epsilon_{m})T/2} \left(\int_{-T/2}^{T/2} e^{i (\epsilon_{n}-\epsilon_{m})t}\langle \chi_{n}(t)|\cos \theta|\chi_{m}(t)\rangle dt\right)
\left(\frac{\sum_{l=0}^{l_{\rm max}} e^{i\left(\epsilon_{n}-\epsilon_{m}\right) l T} }{(l_{\rm max}+1)T}\right) \;.
\label{promedio}
\end{equation}

\noindent
This expression goes to zero when $l_{\rm max}\rightarrow \infty$, since

\begin{equation}
\sum_{l=0}^{\infty} e^{i\left(\epsilon_{n}-\epsilon_{m}\right) l T}=\frac{1}{1-e^{i \left(\epsilon_{n}-\epsilon_{m}\right) T}}  \;.
\label{promedio3}
\end{equation}

There is an exception to this behavior. When the expansion of $\Psi(t)$ contains two degenerate Floquet eigenstates, the average orientation may be nonzero even in the long-time limit since the exponentials are cancelled.  Then, the average is zero only for the two special cases given in Eqs. (\ref{resul3}), and (\ref{resulb3}). Note that, if the initial state is a rotational eigenstate, $\Psi(-T/2)\equiv|J, M\rangle=b_{n}\chi_{n}+b_{m}\chi_{m}$; coefficients $b_{n}$, $b_{m}$ obey the restrictions for Eq. (\ref{resul3}) to hold if $\chi_{n}$, $\chi_{m}$ have the same symmetry, or the restrictions for Eq. (\ref{resulb3}) to hold if they have different symmetry. Therefore, the long-time limit average orientation will be zero when the initial states are field-free eigenstates regardless of the existence of degenerate quasienergy eigenstates. 



\noindent

\section{Numerical results}\label{cuatro}

\subsection{Molecular orientation by a single-cycle pulse}\label{cuatroa}

In this section we present calculations of the instantaneous and average orientation as a function of various parameters. The time-evolved wave function can be obtained from Eq. (\ref{eigenpsi}) after diagonalizing the Floquet Hamiltonian in the extended Hilbert space, which gives eigenpairs $\epsilon$ and $\widehat{\chi}$. This Floquet matrix can be very large. A more efficient method consists on dividing  $(t, t_{0})$ into smaller time intervals of $\tau$ duration, which allows writting the propagator $U$ as a product \cite{moiseyev},

\begin{equation}
U(t,t_{0})=\prod_{m=1}^{M} U\left[m \tau+t_{0},(m-1)\tau+t_{0} \right] \;,
\label{pccp1}
\end{equation}

\noindent
where \cite{moiseyev, peskin}

\begin{equation}
U\left[ m\tau+t_{0},(m-1)\tau+t_{0}\right]=\sum_{n=-\infty}^{\infty} e^{2\pi i n (m \tau+t_{0})/T}
\left\{I\delta_{n,0}-i\frac{\tau}{\hbar}\left[F(\theta)\right]_{n,0}
-\frac{1}{2} \frac{\tau^{2}}{\hbar^{2}}\left[F^{2}(\theta)\right]_{n,0}+\cdots \right\} \;.
\label{moiseyeveq}
\end{equation}

\noindent
where $[F(\theta)]_{n,0}$ is the time integral of the matrix representation of the Floquet Hamiltonian in the rotational basis set:

\begin{equation}
\left[F(\theta)\right]_{n,0}=\frac{1}{T}\int_{t_{0}}^{t_{0}+T} e^{-2 \pi i n t/T} F(\theta) dt \;.
\label{ultima}
\end{equation}

Sugny {\it et al.} \cite{zero} showed that a zero-area pulse can produce large instantaneous orientations when its duration is not too short compared to the rotational period, i.e., in an intermediate regime between the sudden and the adiabatic regimes. Our calculations presented in Fig. \ref{figure1} illustrate the conclusions of the symmetry analysis, i.e,  not only instantaneous orientation but a significant average orientation during the time span of the pulse can be achieved in spite of the presence of space-time symmetries. On the other hand, in the sudden regime, a single-cycle pulse does not produce any noticeable instantaneous orientation while longer pulses, well in the adiabatic regime, can produce instantaneous orientation too. 

Instantaneous orientation depends on the field strength.  For the shorter pulses shown in Fig. \ref{figure1}  (sudden regime)  no significant orientation results even for very high intensities. 
With longer pulses (intermediate case) a significant degree of orientation, asymmetric in time, is obtained. The time asymmetry makes the average time orientation to be nonzero. Finally, for the long adiabatic pulse, shown in Fig. \ref{figure1} $a$, the average orientation is zero because $\cos\theta(t)=-\cos \theta (-t)$. In effect, in this case if the initial state is a field-free eigenstate, the  wave function at any time is given by the Stark eigenfunction of the instantaneous Hamiltonian at time $t$. Therefore, in the adiabatic case, Floquet eigenstates, $\chi(-T/2)$, are just rotational eigenstates that evolve through the corresponding Stark eigenstate of the instantaneous time-dependent Hamiltonian. Parameters corresponding to the longest pulse in Fig.  \ref{figure1} are unphysical, as pulses of that time lenght are not so strong. However, the calculation serves to show the characteristics of the adiabatic regime.

Figure \ref{figure2} shows instantaneous orientations for several initial Floquet eigenstate belonging to what we called above the intermediate regime. The initial states for the time propagation were obtained by diagonalizing the matrix $U(T/2,-T/2)$. This figure illustrates the symmetry relationship, Eq. (\ref{rel1}).

\begin{figure}
\centering
\epsfig{file=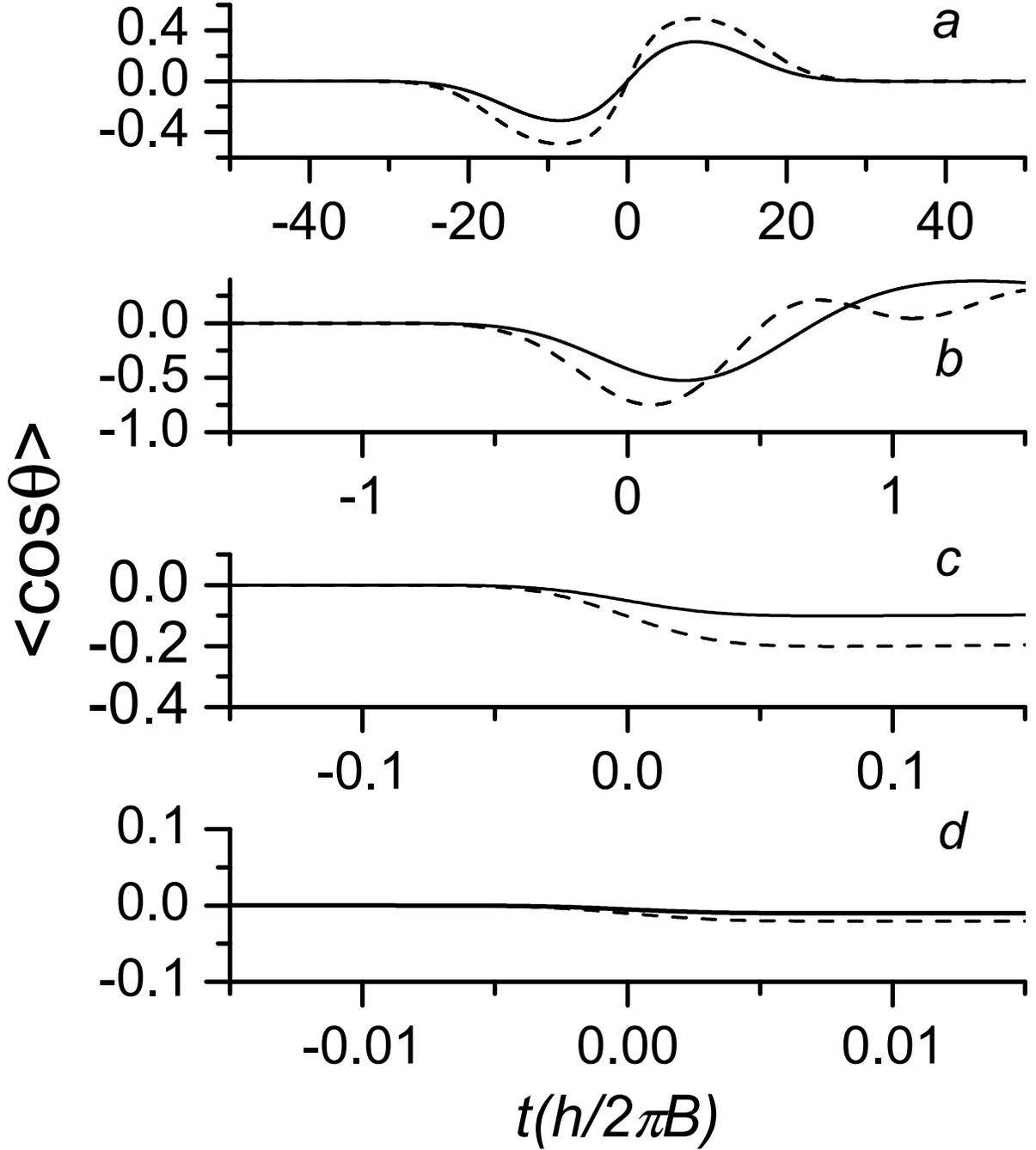,width=\columnwidth}
\caption{Instantaneous orientation, $\langle \cos \theta \rangle$ for initial state, at $t_{0}=-T/2$, $|J=0\rangle$, and eight different single-cycle pulses of zero area. The electric field for each one of these pulses is given by $E(t)=E_{0} e^{-t^{2}/\sigma^{2}}\sin \omega t$. Parameters of the pulses, in reduced units of time $(\hbar/B)$, are: 
($a$) $\sigma=14$, $T=100$, $\omega=0.1$, $\mu E_{0}/B=2$ (solid line), 4 (dashed line), ($b$) $\sigma=0.45$, $T=3$, $\omega=3$, $\mu E_{0}/B=5$ (solid line), 10 (dashed line), ($c$) $\sigma=0.045$, $T=0.3$, $\omega=30$, $\mu E_{0}/B=50$ (solid line), 100 (dashed line), and
($d$) $\sigma=0.0045$, $T=0.03$, $\omega=300$, $\mu E_{0}/B=500$ (solid line), 1000 (dashed line).}
\label{figure1}
\end{figure}

\begin{figure}
\centering
\epsfig{file=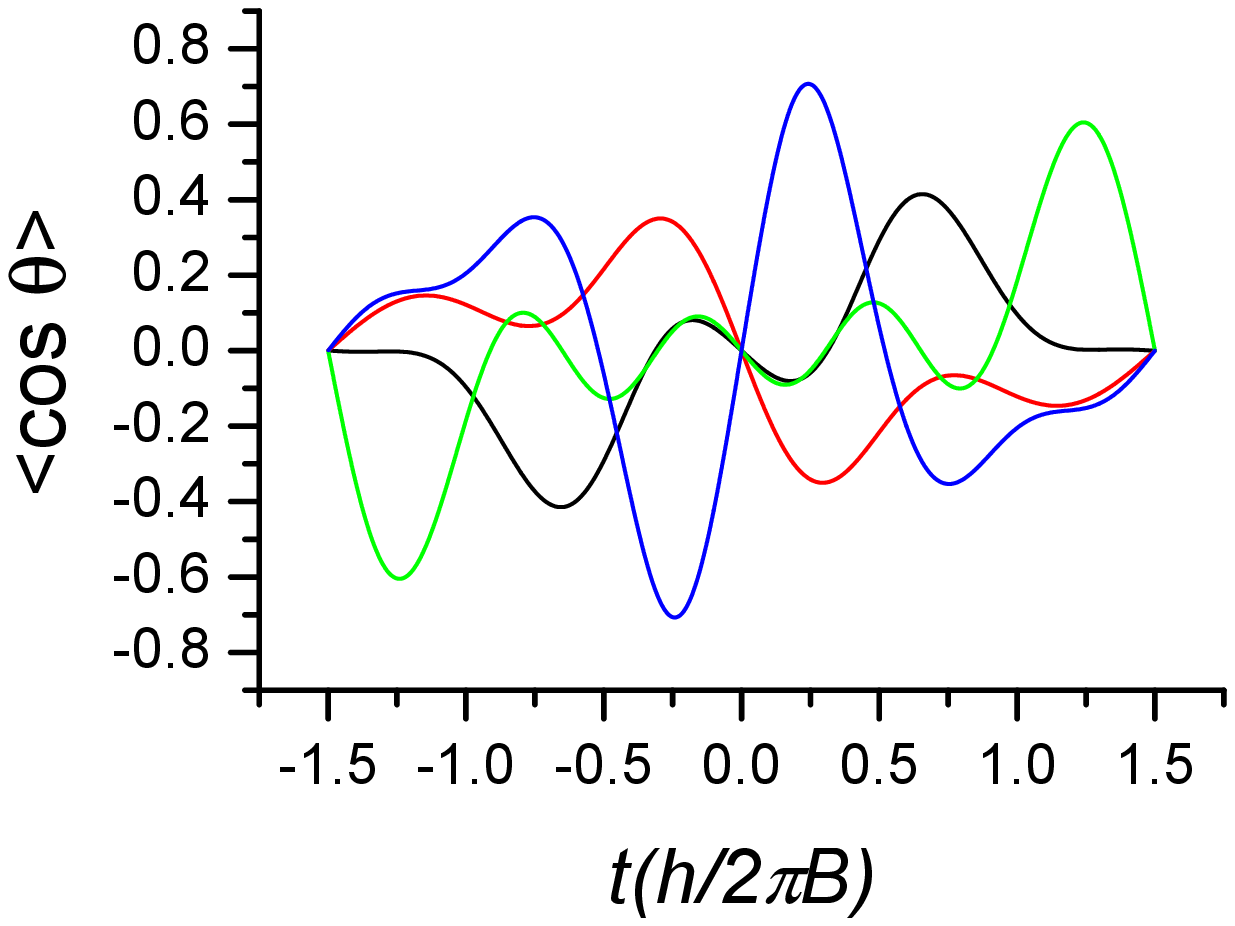,width=\columnwidth}
\caption{Instantaneous orientation, $\langle \cos \theta \rangle$, for initial states corresponding to four different Floquet eigenstates (see Appendix), during a single-cycle pulse. Pulse parameters, in reduced units of time $(\hbar/B)$, are: (a) $\sigma=0.45$, $T=3$, $\omega=3$, $\mu E_{0}/B=10$. The curves correspond to the Floquet eigenstates whose eigenvalues, in $B/\hbar$ units, are: -0.17 (black), -0.58 (red), -0.53 (green), and -0.89 (blue). These wave packets are formed essentially, in the four cases, by combining the five lowest rotational eigenstates. Although large peak orientationes are obtained, the integrated orientation over the pulse duration is strictly zero by symmetry.}
\label{figure2}
\end{figure}

\subsection{Molecular orientation by a sequence of single-cycle pulses}\label{cuatrob} 

Here, we study the results of applying a sequence of identical single-cycle pulses to a molecule in the sudden regime, and in the intermediate regime. It is clear that in the adiabatic regime we would  obtain simply a repetition of what happened for an individual pulse. However, new phenomena appear in the other two regimes. 
Figure \ref{figure3} shows that significant instantaneous orientations are obtained for a sequence of single-cycle pulses in the sudden regime.  For example, the average intrapulse orientation, for the sequence corresponding to $\sigma=0.045$, and pulse number four is -0.32. This should be compared to the average intrapulse orientation in the intermediate case, which for the first pulse is -0.089. In this intermediate regime no further significant increments in average orientation are obtained during successive pulses.  Thus, in the sudden regime, the intrapulse average orientation, with sequences of pulses, is larger than the obtained, in the intermediate regime, with only one pulse. However, the one-period average orientation may be larger for a single pulse of duration comparable to the rotational period than for a sequence whose duration span a rotational period. For example, for $T=\pi$, $\sigma=0.45$, $\mu E_{0}/B=10$, $\omega=3$, the one-period average orientation (corresponding to just one pulse) is -0.089,  for $T=0.1 \pi$, $\sigma=0.045$, $\mu E_{0}/B=100$ $\omega=30$ (ten pulses) is -0.012, and  for $T=0.01 \pi$, $\sigma=0.0045$, $\mu E_{0}/B=1000$, $\omega=300$ (100 pulses) is -0.013. This is not an universal behavior though, as can be seen in Fig. \ref{figure4}. 
This figure shows the average orientation,
during a rotational period, as a function of the field strength, for an initial state $\Psi(-T/2)=|J=0, M=0\rangle$, during one rotational period. Panel $a$ corresponds to a sequence of 10 pulses, and panel $b$ to one pulse.  The maximization of the average orientation during one period is a significant criterion for an efficiency-duration compromise \cite{ben}. Again,  the full duration of each sequency is exactly one rotational period. 

\begin{figure}
\centering
\epsfig{file=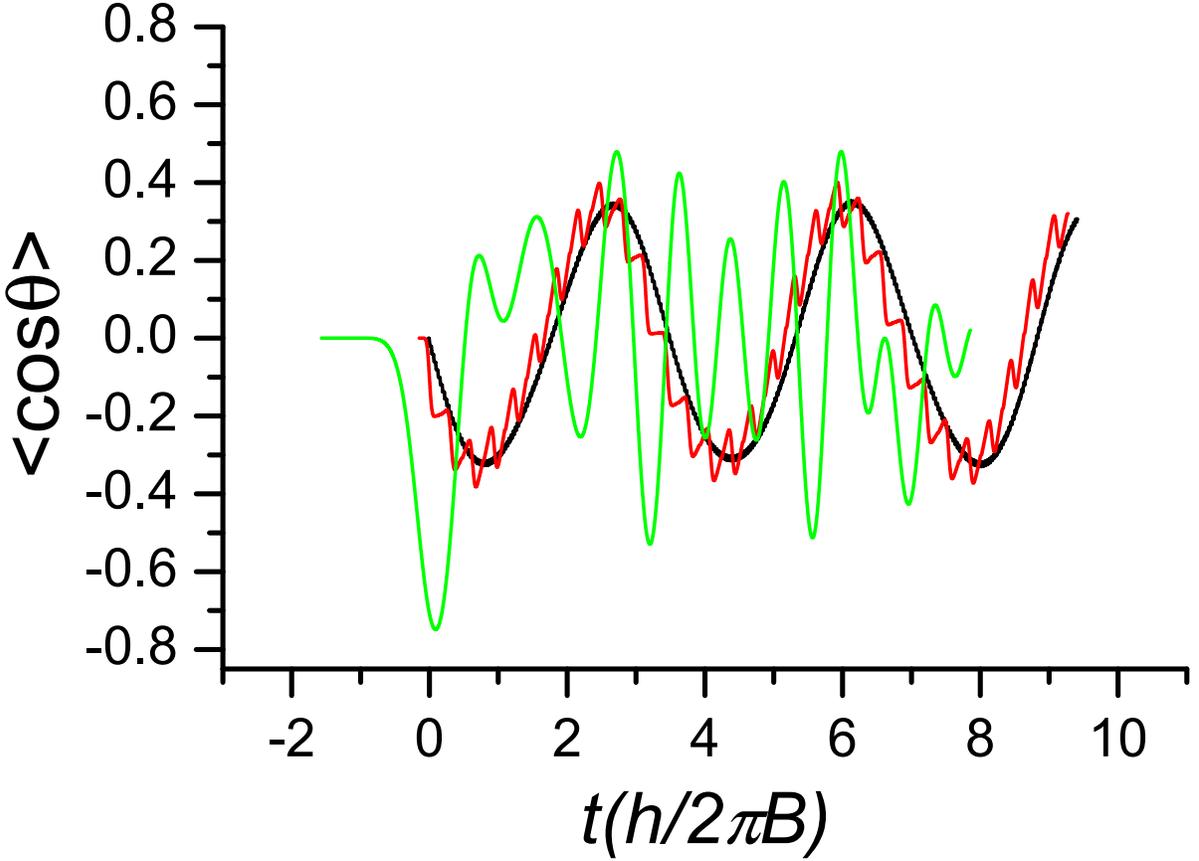,width=\columnwidth}
\caption{Instantaneous orientation, $\langle \cos \theta \rangle$, for three sequences of single-cycle pulses and initial state $\Psi(-T/2)=|J=0\rangle$. Parameters of the pulses, in reduced units of time $(\hbar/B)$, are
$\sigma=0.0045$, $T=0.01\pi$, $\omega=300$, $\mu E_{0}/B=1000$ (black line),  $\sigma=0.045$, $T=0.1 \pi$, $\omega=30$, $\mu E_{0}/B=100$ (red line), and
$\sigma=0.45$, $T=\pi$, $\omega=3$, $\mu E_{0}/B$=10 (green line).}
\label{figure3}
\end{figure}

With current technology, strong anough single-cycle pulses can last no longer than a few ps. Then, a single pulse produces a net average orientation over a time of the order of a rotational period a  only if the rotational constant $B$ is large (for very light molecules). However, the average orientation during a sequence of pulses is controlled by the differences between quasienergies unlike the average orientation over a single pulse, which is controlled by the Floquet eigenvectors \cite{lehman}.  For a sequence of pulses, Eq. (\ref{promedio}) shows that the average orientation is primarily controlled by exponential factors that depend on quasienergy differences multiplied by increasing time,
while for a single pulse the orientation primarily depends on the factors  $\langle \Psi(t)|\cos\theta|\Psi(t)\rangle$ in Eq. (\ref{promedio}), which in turns depend on the composition of the Floquet eigenvectors.

Thus, a nonzero time average can be achieved  by using periodic sequences that span a rotational period as shown in panel $a$ of Fig. \ref{figure4}.  Also, a more stable average 
orientation is obtained with a
sequence of pulses, since the Floquet eigenvalues corresponding to the states that contribute more to instantaneous wave packets remain fairly constant with the field strength. 
On the other hand, for longer pulses frequent avoided crossings exist. At these crossings, the character of the Floquet eigenstates forming the wavepackets may change, giving rise to larger variations on average orientation. 

\begin{figure}
\centering
\epsfig{file=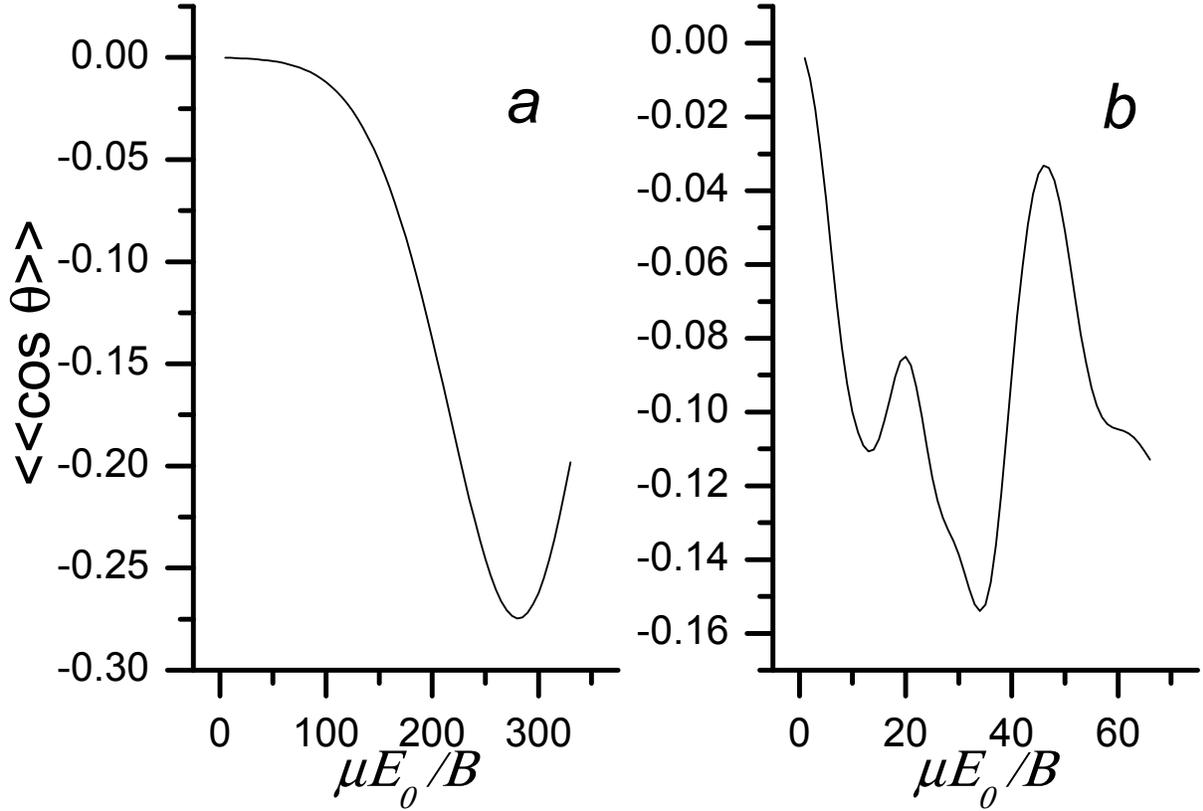,width=\columnwidth}
\caption{Time-average orientation, $\langle\langle \cos \theta \rangle\rangle$, over a rotational period, for two different regimes, as a function of  field strength. Parameters, in reduced units of time $(\hbar/B)$, are: ($a$) $\sigma=0.045$, $T=0.1 \pi$, and ($b$) $\sigma=0.45$, $T=\pi$. Thus, the time average in panel $a$ corresponds, at each point, to ten pulses and  
in panel $b$ to one pulse.}
\label{figure4}
\end{figure}

\section{Discussion} \label{cinco}

Very scarce attention has been paid, in the existing bibliography, to the use of single-cycle pulses of zero area for orienting molecules. Although molecules get sequentially oriented and antioriented as the dipole force breaks rotational parity, the existence of a combined time-space symmetry operation (time-reversal parity \cite{hanggi,lehman}  seems to imply that the average orientation over the full duration of the pulse should be zero. Based on this notion, it is frequently argued that achievement of molecular orientation requires electromagnetic fields asymmetric in time. This is strictly true only in the asymptotic long-time limit. Such a limit is easily reached when high-frequency driving fields are used, but it is not met for low-frequency fields. Nowadays, THz pulses of zero area can be used with exquisite control. Therefore, experiments can be done that use only one pulse or a short sequence of pulses for which the asymptotic limit is far away. We have shown that significant nonzero molecular orientation can be obtained with such fields. The mechanism at work is the existence of nonzero matrix elements for the operator $\cos \theta$ between pairs of Floquet states. These elements enter, multiplied by phase factors, into the expression for the average cosine. This causes the average orientation for a long sequence of pulses to remain zero, as corresponds to the equilibrium or long-time limit. However,  transient breakdown of time-inversion parity leads to significant average orientations for times similar or even slightly longer than a rotational period.

\section*{Acknowledgments} 

Financial support from the Spanish Government through the MICINN project  FIS2010-18799, is acknowledged.

\appendix
\section{Floquet approach}\label{dosa}

Constraints on the average orientation during a pulse or several identical pulses arise as a consequence of the existence of dynamical symmetries in the unitary time propagator, which gives solutions to the time-dependent Schr\"odinger equation as $\Psi(t)=U(t,t_{0})\Psi(t_{0})$. These symmetries can be analyzed more conveniently by considering that the single pulse or the finite sequence of pulses are a member of a infinite periodic train. For a periodic Hamiltonian $H(t+T)=H(t)$ the propagator, $U(t,t_{0})$, satisfies

\begin{equation}
U(t+T,t_{0}+T)=U(t,t_{0}), \; \forall \; t, t_{0} \;.
\label{eq1}
\end{equation}

\noindent
Operators $U(t+T,t)$ are called Floquet operators, and their spectral properties are related to those of the Floquet Hamiltonian,
${\cal F}(t)=H(t)-i\partial/\partial t$. Specifically, eigenfunctions, $\chi$, of ${\cal F}$ are related to those of $U(t_{0}+T,t_{0})$, $\psi$, as follows \cite{yajima}. If 

\begin{equation}
U(t_{0}+T,t_{0}) \psi=e^{-i \epsilon T} \psi\;,
\label{yajim0}
\end{equation}

\noindent
Then, for any $n$,

\begin{equation}
\chi(t)= e^{i (\epsilon+n) (t-t_{0}) } U(t,t_{0}) \psi \;,
\label{relation}
\end{equation}

\noindent
is an eigenfunction of ${\cal F}$ with eigenvalue $\epsilon+n$.

The Floquet Hamiltonian acts in an enlarged Hilbert space where time, designated by $t'$, is treated like a spatial coordinate \cite{peskin, shirley, sambe}. This operator may have no normalizable eigenvectors if its spectrum is continuous. However, heuristic evidence has been given that the Floquet Hamiltonian for a rotating molecule interacting with external fields has a pure point spectrum \cite{ortidense}. Eigenfunctions,  $\widehat{\chi}$, of ${\cal F}$ in this space, can be expanded in terms of a Fourier time basis and spatial basis functions $\phi_{k}(x)$

\begin{equation}
\widehat{\chi}(x,t')=\sum_{k}\left(\sum_{n} d_{kn} e^{2 \pi i n t'/T} \right)\phi_{k}(x) \;.
\label{eq2}
\end{equation}

\noindent
Functions $\chi(t)$, Eq. (\ref{relation}), can be obtained from Eq. (\ref{eq2}) by taking the projection $t'=t$, which gives

\begin{equation}
\chi(x, t)=\sum_{k} c_{k}(t) \phi_{k}(x) \;,
\label{eq34}
\end{equation}

\noindent
where

\begin{equation}
c_{k}(t)=\sum_{n} d_{kn} e^{2\pi i n t/T} \;.
\label{eq4}
\end{equation}

The time-evolved wave function, for an initial $\Psi(t_{0})$, can be expanded as

\begin{equation}
\Psi(t)=\sum_{k} e^{-i \epsilon_{k}(t-t_{0})} \langle \chi_{k}(t_{0})|\Psi(t_{0})\rangle |\chi_{k}(t)\rangle \;.
\label{eigenpsi}
\end{equation}

\noindent
The Floquet Hamiltonian (in reduced units, $B/\hbar$ \cite{ortigoso}) for a rotating linear molecule in the presence of a periodic train of linearly polarized  pulses, and repetition period $T$ is \cite{ortiprl}

\begin{equation}
{\cal F}(t')={\bf{\rm  J}}^{2}-\frac{\mu E_{0}}{B} \exp(-t'^{2}/\sigma^{2}) \sin(\omega t'+\phi) \cos\theta - i\frac{\hbar}{B} \frac{\partial}{\partial t'} \;,
\label{floqueth}
\end{equation}

\noindent
where $\omega$ is the carrier frequency,  ${\bf{\rm J}}$ is the angular momentum vector, $B$ the rotational constant, $\mu$  the permanent dipole moment, $\theta$ is the angle between the polarization vector of the field and  the internuclear axis, and $\sigma\approx 3\delta/5$ where $\delta$ is the Gaussian half-width at half-maximum. The carrier envelope phase, $\phi$, is zero for single-cycle pulses of zero area. For this particular operator, Eq. (\ref{eq2}) can be expanded as a summation of rotational eigenfunctions,
$\langle\theta, \Phi|J,M\rangle$, with $M$ the projection of the $\bf{\rm J}$ vector along the polarization direction of the electric field, and $\Phi$ the other Eulerian angles that define the direction of ${\bf {\rm J}}$. In this Article, the summation index $k$, in Eq. (\ref{eq2}), is substituted by $J$.


\begin{thebibliography}{}


\bibitem{loesch} H. J. Loesch, Ann. Rev. Phys. Chem. {\bf 46}, 555 (1995).

\bibitem{friedrich1} B. Friedrich and D. Herschbach, Z. Phys. D {\bf 24}, 25 (1992).

\bibitem{seide1} H. Stapelfeldt and T. Seideman, Rev. Mod. Phys. {\bf 75}, 543 (2003).

\bibitem{seide1b} T. Seideman and E. Hamilton, Adv. At., Mol., Opt. Phys. {\bf 52}, 289 (2005).

\bibitem{bretis2} B. Friedrich and D. R. Herschbach, Phys. Rev. Lett. {\bf  74}, 4623 (1995).

\bibitem{japo} H. Sakai, S. Minemoto, H. Nanjo, H. Tanji, and T. Suzuki, Phys. Rev. Lett. {\bf 90}, 083001 (2003).

\bibitem{bretis} B. Friedrich and D. Herschbach, J. Chem. Phys. {\bf 111}, 6157, (1999);  J. Phys. Chem. A {\bf 103}, 10280 (1999).

\bibitem{bretism} L. Cai, J. Marango, and B. Friedrich, Phys. Rev. Lett. {\bf 86}, 775 (2001).

\bibitem{bretisb}  S Minemoto, H. Nanjo, H. Tanji, T. Suzuki, and H. Sakai, J. Chem. Phys. {\bf 118}, 4052 (2003).

\bibitem{ortisantos} J. Ortigoso and J. Santos, Phys. Rev. A {\bf 72}, 053401 (2005).

\bibitem{feres} J. H. Nielsen, H. Stapelfeldt, J. K\"upper, B. Friedrich, J. J. Omiste, and R. Gonz\'alez-F\'erez, Phys. Rev. Lett. {\bf 108}, 193001 (2012).

\bibitem{twocolor} C. M. Dion, A. D. Bandrauk, O. Atabek, A. Keller, H. Umeda, and Y. Fujimura, Chem. Phys. Lett. {\bf 302}, 215 (1999); S. Gu\'erin, L. P. Yatsenko, H. R. Jauslin, O. Faucher, and B. Lavorel, Phys. Rev. Lett. {\bf 88}, 233601 (2002).

\bibitem{twocolor2} T. Kanai and H. Sakai, J. Chem. Phys. {\bf 115}, 5492 (2001); S. De, I. Znakovskaya, D. Ray, F. Anis, N. G. Johnson, I. A. Bocharova, M. Magrakvelidze, B. D. Wsry, C. L. Cocke, I. V. Litvinyuk, and M. F. Kling, 
Phys. Rev. Lett. {\bf 103}, 153002 (2009).

\bibitem{twocolor2b} R. Tehini and D. Sugny, Phys. Rev. A  {\bf 77}, 023407 (2008).

\bibitem{chin} Y. X. Huang, Y. J. Yang, B. Wu, F. M. Guo, and Q. R. Zhu, Chin. Phys. Lett. {\bf 25}, 1259 (2008).

\bibitem{fcp} W. H Hu, C. C. Shu, Y. C. Han, K. J. Yuan, and S. L. Cong, Chem. Phys. Lett. {\bf 480}, 193 (2009); C. C. Shu, K. J. Yuan, W. H. Hu, and S. L. Cong, Phys. Rev. A {\bf 80}, 011401(R) (2009); J. Chem. Phys., {\bf 132}, 244311 (2010).

\bibitem{orti1} J. Ortigoso, Phys. Rev. A {\bf 57}, 4592 (1998). 

\bibitem{salomon} J. Salomon, C. M. Dion, and G. Turicini, J. Chem. Phys. {\bf 123}, 144310 (2005).

\bibitem{stapel1} H. Stapelfeldt, Phys. Scrip. {\bf T110}, 132 (2004).

\bibitem{japo2}  A. Goban, S. Minemoto, and H. Sakai, {\it Phys. Rev. Lett.}, {\bf 101}, 013001 (2008);
K. Oda, M. Hita, S. Minemoto, and H. Sakai, {\it ibid}, {\bf 104}, 213901 (2010).


\bibitem{orien1} N. E. Henriksen, Chem. Phys. Lett. {\bf 312}, 196 (1999); C. M. Dion, A. Keller and O. Atabek, Eur. Phys. J. D {\bf 14}, 249 (2001);  M. Machholm and N. E. Henriksen, Phys. Rev. Lett. {\bf 87}, 193001 (2001); D. Sugny, A. Keller, O. Atabek, D. Daems, C. M. Dion, S. Guerin, and H.R. Jauslin, Phys. Rev. A {\bf 69}, 033402 (2004).

\bibitem{hanggi} S. Denisov, L. Morales-Molina, S. Flach, and P. H\"anggi, Phys. Rev. A {\bf 75}, 063424 (2007).

\bibitem{lehman}  J. Lehmann, S. Kohler, P. H\"anggi, and A. Nitzan, Phys. Rev. Lett. {\bf 88}, 228305 (2002);  J. Chem. Phys. {\bf 118}, 3283 (2003).

\bibitem{santos} D. Poletti, G. Benenti, G. Casati, P. H\"anggi, and B. Li, Phys. Rev. Lett. {\bf 102}, 130604 (2009).

\bibitem{santosb}  J. Santos, R. A. Molina, J. Ortigoso, and M. Rodr\'iguez, Phys. Rev. A {\bf 84}, 023614 (2011).

\bibitem{salger} T. Salger, S. Kling, T. Hecking, C. Geckeler, L. Morales-Molina, and M. Weitz, Science {\bf 326}, 1241 (2009).

\bibitem{zero} D. Sugny, A. Keller, O. Atabek, D. Daems, S. Guerin, and H.R. Jauslin, Phys. Rev. A {\bf 69}, 043407 (2004).

\bibitem{bob} S. Fleischer, Y. Zhou, R. W. Field, and K. A. Nelson, Phys. Rev. Lett. {\bf 107}, 163603 (2011).

\bibitem{yajima} K. Yajima, J. Math. Soc. Jpn. {\bf 29}, 729 (1977).

\bibitem{bunker} P. R. Bunker, {\it Molecular Symmetry and Spectroscopy}, Academic Press, Inc., New York, 1979.

\bibitem{moiseyev} N. Moiseyev, Comments At. Mol. Phys. {\bf 31}, 87 (1995).

\bibitem{peskin} U. Peskin and N. Moiseyev, J. Chem. Phys. {\bf 99}, 4590 (1993).

\bibitem{ben} A. Ben Haj-Yedder, A. Auger, C. M. Dion, E. Cances, A. Keller, C. Le Bris, and O. Atabek, Phys. Rev. A {\bf 66 }, 063401 (2002).

\bibitem{shirley} J. H. Shirley, Phys. Rev. {\bf 138}, B979 (1965).

\bibitem{sambe} H. Sambe, Phys. Rev. A  {\bf 7}, 2203 (1973).

\bibitem{ortidense} J. Ortigoso, M. Rodr\'iguez, J. Santos, A. Karpati, and V. Szalay,  J. Chem. Phys. {\bf 132}, 074105 (2010).

\bibitem{ortigoso} J. Ortigoso, M. Rodr\'iguez, M. Gupta, and B. Friedrich, J. Chem. Phys. {\bf 110}, 3870 (1999). 

\bibitem{ortiprl} J. Ortigoso, Phys. Rev. Lett. {\bf 93}, 073001 (2004).
\end{thebibliography}
\end{document}